\documentclass[10pt,pre,twocolumn]{revtex4-1}

\usepackage{dcolumn}						
\usepackage{bm}							

\usepackage{graphicx} 
\usepackage{amssymb}
\usepackage{latexsym}
\usepackage{hyperref}
\usepackage{color}

\newcommand \be {\begin{equation}}
\newcommand \ee {\end{equation}}
\newcommand \bea {\begin{eqnarray}}
\newcommand \eea {\end{eqnarray}}

\usepackage{amsmath}

\begin{document}

\title{Dynamics of the bouncing ball}
\author{J.-Y. Chastaing, E. Bertin$^{\dagger}$ and J.-C. G\'eminard}
\email{jean-christophe.geminard@ens-lyon.fr}
\affiliation{Laboratoire de Physique, Ecole Normale Sup\'erieure de Lyon,
CNRS, Universit\'e de Lyon, 46 All\'ee d\'\,Italie, 69364 Lyon Cedex, France \\
$^{\dagger}$ Laboratoire Interdisciplinaire de Physique (LIPhy), 140 Avenue de la Physique - BP 87, 38402 Saint Martin d'Hères, France}
\date{\today}

\begin{abstract}

We describe an experiment dedicated to the study of the trajectories of a ball bouncing randomly on a vibrating plate.
The system was originally used, considering a sinusoidal vibration, to illustrate period doubling and the route to chaos.
Our experimental device makes it possible to impose, to the plate, arbitrary trajectories and not only sinusoidal or random, as is generally the case.
We show that the entire trajectory of the ball can still be reconstructed from the measurement of the collisions times.
First, we make use of the experimental system to introduce the notion of dissipative collisions and to propose three different ways to measure the associated restitution coefficient. Then, we report on correlations in the chaotic regime and discuss theoretically the complex pattern which they exhibit in the case of a sinusoidal vibration. At last, we show that the use of an aperiodic motion makes it possible to get rid of part of the correlations and to discuss theoretically the average energy of the ball in the chaotic regime.

\end{abstract}

\maketitle

\section{Introduction}

Let a ball fall vertically onto a horizontal floor.
One knows that the ball bounces repetitively before coming to rest. Indeed, the collisions between the ball and the floor are dissipative,
\textit{i.e.} part of the energy is dissipated during the collisions.
As a consequence the maximum bouncing height after a collision is less than the maximum height previously achieved.
Even more, assuming that the energy lost at each of the collisions is a constant fraction of the energy of the ball,
one can predict that the ball bounces an infinite number of times in a finite time.
This {\it collisional singularity} is a problem introduced in classical mechanics courses  and, thus, in textbooks.

A way to maintain the ball in permanent motion is to move periodically
the floor up and down as one would do juggling with a racket and a ball, by moving alternatively the racket up and down with the appropriate frequency and amplitude.
It has been achieved in many, more or less expensive and delicate, experimental configurations and proved to  be a very interesting model for the study, in particular, of the transition to chaos.\cite{Pieranski1983,Pieranski1985}

In the present article, we describe a simple realization of the experiment:
a ball is bouncing vertically above a vibrated plate.
Our experimental techniques, based on the measurement of the collision times,
add to previous studies the possibility to consider arbitrary motion of the plate. 
It makes it possible to introduce, with educational aims, basic concepts of the problem
and to answer several basic questions such as, for instance, ``How can we assess reliable values of the fraction of energy lost at each collision?'',
or ``What is the average energy of the ball?''.

\section{Experimental principle and setup}
\label{sec:basic_concepts}

\subsection{Experimental principle}
\label{sec:principle}

\begin{figure}[b]
\begin{center}
\includegraphics[width=.9\columnwidth]{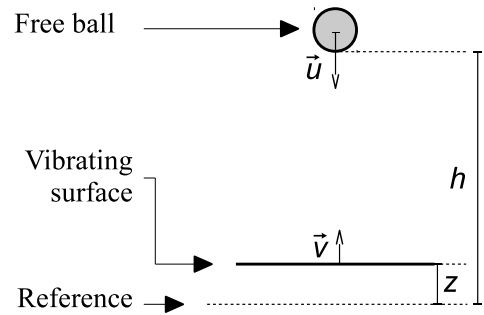}
\end{center}\caption{\label{fig:principle} Sketch of the experimental principle.}
\end{figure}

The experiment is very simple (Fig.~\ref{fig:principle}):
a horizontal plate is vibrated vertically and a ball is bouncing freely above it.\cite{Pieranski1983,Tufillaro1986}
The motion of the plate is characterized by its altitude $z(t)$ and velocity $v(t) \equiv dz/dt$ at time $t$.
In the same way, the motion of the ball is characterized by its altitude $h(t)$ and velocity $u(t) \equiv dh/dt$.

During its motion, the ball repeatedly falls freely, collides with the vibrating plate and bounces back after the collision. 
The overall trajectory of the ball consists of a series of collisions, at time $t_\mathrm{n}$ ($\mathrm n$ is the index of the collision),
separated by a time $\Delta t_\mathrm{n} \equiv t_\mathrm{n+1}-t_\mathrm{n}$ of free fall.

\subsubsection{Collision and restitution coefficient}
\label{sec:restitution_def}

Assuming that the collisions are instantaneous, one can write a simple relation
between the velocity of the ball before, $u_\mathrm{n}^\mathrm{-}$, and after, $u_\mathrm{n}^\mathrm{+}$,
the collision \cite{Bernstein1977}:
\begin{equation}
u_\mathrm{n}^\mathrm{+} - v_\mathrm{n} = - e\,( u_\mathrm{n}^\mathrm{-} - v_\mathrm{n} )
\label{eq:restitution}
\end{equation}
where $v_\mathrm{n}$ stands for the velocity of the plate at $t_\mathrm{n}$.
The coefficient $e$ is, by definition, the {\it restitution coefficient} which accounts for the energy loss associated with the event. 
In good approximation for most of the practical cases, $e$ is independent from the impact velocity $u_\mathrm{n}^\mathrm{-} - v_\mathrm{n}$.\cite{Johnson1985,Tillett1954,Reed1985,Sondergaard1990}
It depends only on the two colliding objects and ranges from zero for completely inelastic collisions to the unity for elastic collisions.

\subsubsection{Free fall}
\label{sec:flight}

Between two subsequent collisions, the ball is only subjected to gravity, $g$.
Thus, one can write, from $t_\mathrm{n}$ to $t_\mathrm{n+1}$, the altitude of the ball as a function of time in the form:
\begin{equation}
h(t) = h_\mathrm{n} + u_\mathrm{n}^\mathrm{+} (t - t_\mathrm{n}) - \frac{g}{2} (t - t_\mathrm{n})^2.
\label{altitude}
\end{equation}
where $h_\mathrm{n} \equiv h(t_\mathrm{n})$.
From Eq.~(\ref{altitude}), one can determine the next collision time $t_\mathrm{n+1}$ by considering that
$h(t_\mathrm{n+1}) = z(t_\mathrm{n+1})$ and, then, the velocity of the ball before the collision:
\begin{equation}
u_\mathrm{n+1}^\mathrm{-} = u_\mathrm{n}^\mathrm{+} - g (t_\mathrm{n+1} - t_\mathrm{n}).
\label{next_velocity}
\end{equation}

\subsubsection{Ball trajectory}
\label{sec:trajectory}

The dynamics of the bouncing ball is governed by Eqs.~(\ref{eq:restitution}) to (\ref{next_velocity}),
together with $h(t_\mathrm{n+1}) = z(t_\mathrm{n+1})$.
It is particularly interesting that the knowledge
of the impact times $t_\mathrm{n}$ makes it possible to reconstruct the entire trajectory $h(t)$
of the bouncing ball for any (known) plate trajectory $z(t)$.
Indeed, considering Eq.~(\ref{eq:restitution}) and noticing that $h_\mathrm{n} = z(t_\mathrm{n})$,
one easily obtains $u_\mathrm{n}^\mathrm{+}$ from the times $t_\mathrm{n}$ and $t_\mathrm{n+1}$ and
altitudes $z(t_\mathrm{n})$ and $z(t_\mathrm{n+1})$. 
Thus, the principle of the experiment will be to impose the plate motion $z(t)$
and to determine the impact times $t_\mathrm{n}$ from which the entire trajectory
of the ball will be reconstructed and analyzed.

\subsection{Experimental setup and techniques}
\label{sec:setup}

\subsubsection{Experimental device}
\label{sec:device}

The literature reports several practical realizations of the experimental principle.\cite{Pieranski1983,Tufillaro1986}
In the most simple version, the vibrating surface is a concave lens driven by a loudspeaker.
The latter is attached to a table top so that its membrane vibrates vertically.
The use of a concave lens of large radius of curvature avoids that the ball escape the system due to any, even very small, tilt of the plate.

\begin{figure}[t]
\begin{center}
\includegraphics[width=.9\columnwidth]{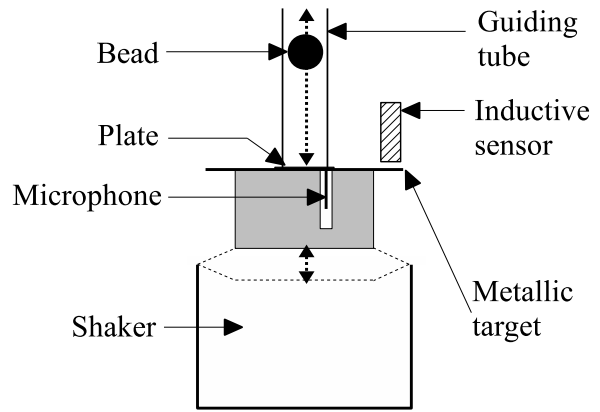}
\end{center}
\caption{\label{fig:setup} Sketch of the experimental setup.}
\end{figure}

Our experimental system, which differs slightly from these configurations, is adapted from a former experimental study.\cite{Geminard2003}
An permanent magnet shaker (Br\"uel\&Kjaer, Type 4803) imposes the vertical displacement of a microscope glass slide (Fig.~\ref{fig:setup}).
The bouncing ball is a steel bead of diameter 8~mm. In order to avoid that it escapes the system, the latter is guided by a tube placed along the vertical axis. 
The play (typically 0.1~mm) between the inner surface and the bead insures that the friction is negligible while avoiding significant lateral motion.

In addition, the use of a powerful shaker makes it possible to increase the inertia of the system by adding a large mass (800~g) below the plate, which avoids any significant receding of the plate due to the impact with the bead.
The ballast also increases the vertical distance between the plate and the magnetic shaker, which avoids that the magnetic field alter the dynamics of the bead.

The power amplifier (Kepco, BOP50-4M) driving the shaker receives the signal from the analog output of a data acquisition board (National Instruments, Model 6251).
Thus, one can impose an arbitrary trajectory to the plate, and not only sinusoidal as usual.
One measures its vertical position $z(t)$ by means of a non-contact inductive sensor
(Electrocorp, EMD1053, sensing range 0-3 mm, resolution 1 $\mu$m) whose signal is recorded by one analog input of the same acquisition board.
The sensor, immobile, measures the distance between its head and a metallic target attached to the plate mount.
The typical frequency of the vibration used in the present study will range from 20 to 80~Hz,
for a resulting acceleration up to 3~$g$. The typical vertical displacement is thus of the order of 0.1 to 2~mm.

Several authors proposed to detect the collisions by recording the associated acoustic emission.\cite{Bernstein1977,Smith1981,Stensgaard2001,Geminard2003}
Following this idea, we use a pinducer microphone attached to the glass slide mount, its head in contact with the lower glass surface.
The signal is sent to a second input of the acquisition board and recorded,
together with the position of the plate, during the whole experimental time.
The acquisition rate, $f_\mathrm{a}=5$~kHz, insures a sufficient temporal resolution without increasing too much the number of data points.

\begin{figure}[t]
\begin{center}
\includegraphics[width=.8\columnwidth]{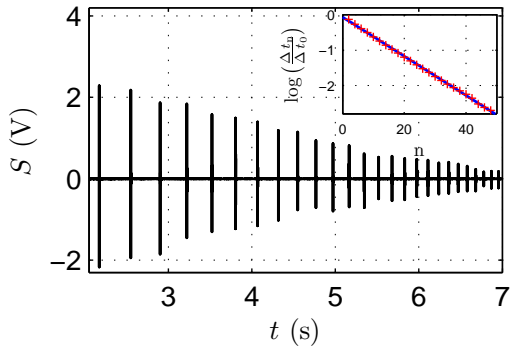}
\end{center}
\caption{\label{fig:damping} 
Signal $S(t)$ from the microphone for a ball bouncing on the plate at rest. 
Inset: Logarithm of the normalized flight time $\log \left( \Delta t_\mathrm{n}/\Delta t_\mathrm{0} \right)$ vs. collision index $\mathrm{n}$ -- One observes a nice linear behavior, thus an exponential decay of $\Delta t_\mathrm{n}$ with n.}
\end{figure}

\begin{figure}[t]
\begin{center}
\includegraphics[width=.8\columnwidth]{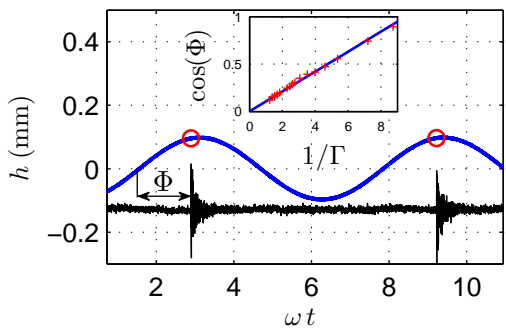}
\end{center}
\caption{\label{fig:periodic}
Altitude $h(t)$ (thick blue line) vs. time $t$ -- The signal from the microphone $S(t)$ is reported in arbitrary units to show the collisions (circles).
The ball, in spite of some fluctuations, collides with the plate once per cycle for the same phase $\Phi$~[$\Gamma = 0.25$, $f = 25$~Hz].
Inset: $\cos(\Phi)$ vs. $1/\Gamma$ -- The size of the markers accounts for the error [$f = 25$~Hz].}
\end{figure}

\subsubsection{Data analysis}
\label{sec:data analysis}

The data from the acquisition board are subsequently analyzed (MathWorks, Matlab R2010b) in order to extract 
the impact times $\{ t_\mathrm{n} \}$, the plate positions $\{ z_\mathrm{n} \}$ and the velocities $\{ v_\mathrm{n} \}$.
Typical examples are reported in Figs.~\ref{fig:damping}~and~\ref{fig:periodic}.

The procedure is as follows. For each collision, the signal from the microphone exhibits a large peak and
subsequent oscillations. Indeed, the collision, which is very short, excites vibrations of the glass slide.
Since the vibrations frequency is higher than the acquisition rate $f_\mathrm{a}=5$~kHz, we assume the collision
time $t_\mathrm{n}$ to be $t_\mathrm{n} = t_p - 1/(2 f_\mathrm{a})$, where $t_p$ is the time of the largest peak.
Then, one obtains the corresponding vertical position, $z_\mathrm{n}$ of the surface,
and thus the altitude of the ball $h_\mathrm{n}$, from the signal from the inductive sensor.The velocity $v_\mathrm{n}$ is obtained
by interpolating $z(t)$ by a parabola over 50 points around the collision and by calculating the slope at $t_\mathrm{n}$. 
As a result, the temporal resolution of the measurements is thus about the inverse of the acquisition rate, $\frac{1}{f_\mathrm{a}}=0.2$ ms,
thus 0.4\% to 1.6\% of the vibration period. The accuracy in the plate position is mainly limited by the resolution of the inductive sensor
(typically 1~$\mu$m), again 0.5\% to 1\% of the typical amplitude of the vibration.

\section{Experimental results}
\label{sec:results}

Using a periodic vibration, we first describe three different methods to measure the restitution coefficient~$e$ which is the single physical parameter of the problem (Sec.~\ref{sec:restitution}).
Two methods are classical. The third one
is simple in its principle but original and more difficult to achieve experimentally.
It consists in verifying the collision rule (Eq.~\ref{eq:restitution}) at each collision.
Then, we show that, even in the chaotic regime reached for intense vibrations, the motions of the ball and of the plate are strongly correlated: the velocity of the plate seen, in average, by the falling ball depends on its fall velocity (Sec.~\ref{sec:correlations_S}).
We discuss, in the same section, the physical origin of the correlations and propose a model to describe the complex pattern revealed experimentally.
Using an aperiodic, but well-controlled, vibration (\ref{sec:platemotion}), which is made possible by our original device, we describe how to get rid, at least partially, of the ball-plate correlations (Sec.~\ref{sec:correlations_FM}).
At last, in this context, we measure, and then account theoretically for, the average energy of the ball in the chaotic regime (\ref{sec:energy of the bead}).

\subsection{Periodic vibrations}

\subsubsection{Collisions and restitution coefficient}
\label{sec:restitution}

Let us consider first that the plate does not vibrate.
One observes that the height reached after each bounce decreases until the ball comes to rest.\cite{Falcon1998,Bernstein1977}
By considering the collision law (Eq.~\ref{eq:restitution}) with $v_\mathrm{n} = 0$ and the free flight,
which leads to $u_\mathrm{n+1}^\mathrm{-} = - u_\mathrm{n}^\mathrm{+}$, one obtains that $u_\mathrm{n}^\mathrm{-}$
decreases exponentially with n. Accordingly, the duration of the flights obeys:
\begin{equation}
\Delta t_\mathrm{n} = \Delta t_\mathrm{0}\,e^\mathrm{n}  = \Delta t_\mathrm{0}\,\exp{(- \mathrm{n} / \mathrm{n}_\mathrm{c})}.
\label{eq:time series}
\end{equation}
with $\mathrm{n}_\mathrm{c} \equiv -1/\log{e}$ (We remind ourselves here that $e < 1$ and, thus, that $\log{e} < 0$).
We report in Fig.~\ref{fig:damping} the signal from the microphone when the ball
is released from a height of about 15~cm above the plate at rest.
We observe, indeed, a nice exponential decay of the duration between the collisions, $\Delta t_\mathrm{n}$, and the interpolation of the experimental data leads to $e = 0.946 \pm 0.005$.

Let us now consider a sinusoidal vibration, $z(t) = A \cos{(\omega t)}$, where $A$ is the vibration amplitude
and $\omega$ its pulsation (we denote $f=\frac{\omega}{2\,\pi}$ the frequency of vibration).\cite{Tufillaro1986}
Depending on $A$, $\omega$ and on the initial conditions, the ball can exhibit a periodic trajectory,
in particular bouncing once per cycle of the plate motion.
The trajectory being periodic, none of its characteristics depends on n.
In particular, $h_\mathrm{n+1} = h_\mathrm{n}$ so that, from Eq.~(\ref{altitude}), $u_\mathrm{n+1}^\mathrm{-} = - u_\mathrm{n}^\mathrm{+}$
and, finally, $u_\mathrm{n}^\mathrm{-} = - u_\mathrm{n}^\mathrm{+}$.
Moreover, the duration of the free fall is equal to the period of the plate vibration, $\Delta t = \frac{2\,u_\mathrm{n}^\mathrm{+}}{g} = \frac{2\pi}{\omega}$,
which leads to $u_\mathrm{n}^\mathrm{+} = \frac{g \pi}{\omega}$.
Finally, the collision law (Eq.~\ref{eq:restitution}) imposes the plate velocity at the impact,
$v_\mathrm{n} = \frac{1-e}{1+e} \frac{g \pi}{\omega}$.
A convenient way to write the condition is to consider the phase $\Phi = \omega t_\mathrm{n}~~[2\pi]$, which in principle does not depend on $\mathrm{n}$,
and to write, from the plate trajectory $z(t)$, $v_\mathrm{n} = A\,\omega\,\cos{(\Phi)}$.
One gets:
\begin{equation}
\cos{(\Phi)} = k\,\frac{1-e}{1+e}\,\frac{\pi}{\Gamma}  
\label{eq:phase}
\end{equation}
where $\Gamma \equiv A \omega^2 / g$ is the reduced plate acceleration.
Experimentally, such periodic trajectory is achieved provided one makes a pertinent choice of the vibration characteristics.\cite{Tufillaro1986}
On the one hand, the reduced acceleration $\Gamma$ must be large enough to insure that Eq.~(\ref{eq:phase}) has a solution, i.e. $\frac{1-e}{1+e}\,\frac{\pi}{\Gamma} \le 1$.
This leads to the condition that $\Gamma \ge \frac{1-e}{1+e}\,{\pi}$ (typically $\Gamma \gtrsim 0.1$ with $e = 0.95$).
On the other hand, the reduced acceleration $\Gamma$ must be small enough to avoid period doubling.\cite{Pieranski1983}
We tune $\Gamma$, and study the associated variation of the phase $\Phi$, by varying the amplitude $A$ at a fixed frequency $f = 25$~Hz.
We observe in Fig.~\ref{fig:periodic},
a linear dependence of $\cos{(\Phi)}$ on $1/\Gamma$ from which we get
$e = 0.935 \pm 0.010$.

\begin{figure}[t]
\begin{center}
\includegraphics[width=0.8\columnwidth]{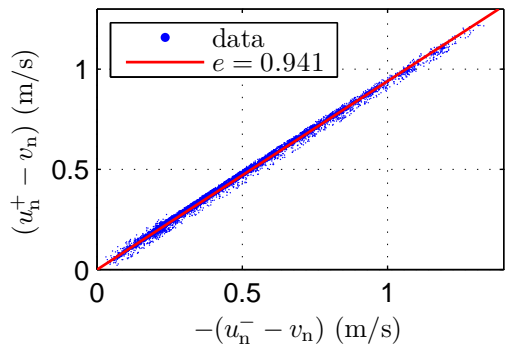}
\end{center}
\caption{\label{fig:chaotic}
Relative outgoing velocity $(u_\mathrm{n}^\mathrm{+} - v_\mathrm{n})$ vs. impact velocity $- (u_\mathrm{n}^\mathrm{-} - v_\mathrm{n})$ -- Dots: experimental data, Red line: best linear fit [$\Gamma = 2.35$, $f = 25$~Hz, $\sim 10^3$ impacts].}
\end{figure}

The most direct way, but the most difficult experimentally, to measure the restitution coefficient is to check, at each impact,
the collision law (Eq.~\ref{eq:restitution}) by displaying the relative outgoing velocity $(u_\mathrm{n}^\mathrm{+} - v_\mathrm{n})$
as a function of the relative incoming velocity $(u_\mathrm{n}^\mathrm{-} - v_\mathrm{n})$.
Experimentally, in order to explore a large range of incoming velocities, we chose a large reduced acceleration ($\Gamma=2.35$) such
that the ball experiences a chaotic trajectory.\cite{Tufillaro1986,Vogel2011,Holmes1982,Luck1993,Oliveira1997,Pieranski1985,Pieranski1983}
We detect a set of collision times, $\{t_\mathrm{n}\}$, from which we deduce $\{h_\mathrm{n}\}$, $\{v_\mathrm{n}\}$, $\{u_\mathrm{n}^\mathrm{-}\}$
and $\{u_\mathrm{n}^\mathrm{+}\}$.
We then confront the collision law (Eq.~\ref{eq:restitution}) with experimental data (Fig.~\ref{fig:chaotic}).
We note that about $5\%$ of the data points are totally out of the collision law (not shown), which mainly occurs for small velocities.
Indeed, the system cannot detect impacts that are occurring too close to one another
because of the finite relaxation time of the plate vibrations.
If one impact is missed, the data corresponding to the impacts immediately
before and after are necessarily wrong. Consequently, one can automatically correct
or simply delete the pairs of those non-physical data. Provided that the erroneous
points are suppressed, the interpolation of the experimental data leads to $e = 0.941 \pm 0.005$.

We proposed three different methods to measure the restitution coefficient $e$.
Notice that the measurements in the periodic regime correspond to a single value of
$- (u_\mathrm{n}^\mathrm{-} - v_\mathrm{n})$ whereas the impact velocity continuously decreases in the first method and is random in the last one. 
However, we did not observe any dependence of the restitution coefficient on the impact velocity, which indicates that, even if such effects exist,\cite{Bernstein1977,Falcon1998} they are small and out-of-reach of our experimental device.
We obtained successively $e = 0.941 \pm 0.005$, $e = 0.935 \pm 0.010$ and $e = 0.946 \pm 0.005$, the three estimates being in quantitative agreement, which proves that all three methods are reliable.

\subsubsection{Ball-Plate correlations}
\label{sec:correlations_S}

Using our simple experimental device, one can also report statistical properties of the ball trajectory in the chaotic regime.
As an example, we chose large reduced accelerations such that the ball explores a large range of incoming velocities in a chaotic way
and report on the correlation between the ball and plate velocities at the collision.

\begin{figure}[t]
\begin{center}
\includegraphics[width=0.8\columnwidth]{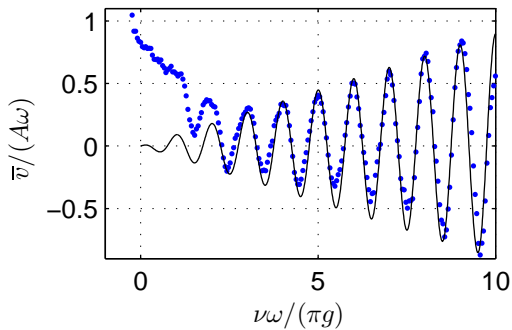}
\end{center}
\caption{\label{fig:va_vp_S} Averaged plate velocity $\overline{v}$ vs. incoming velocity $\nu$ for a sinusoidal vibration --
Dots: experiments with $\Gamma = 2.35$; Line: theoretical prediction with $\Gamma = 35$.}
\end{figure}

To reveal correlations, we report in Fig.~\ref{fig:va_vp_S} the average plate velocity $\overline{v}$ as a function of the incoming velocity $-u^{-}$.
To do so, from a set of data $\{ v_\mathrm{n} \}$ and $\{u_\mathrm{n}^{-}\}$ ($10^3$ impacts), we average the values of $v_\mathrm{n}$ which are associated 
with the same value of (strictly speaking, that are in a given range around) the incoming velocity $\nu \equiv -u_\mathrm{n}^{-}$.
In the absence of correlations, we expect $\overline{v}=0~ (\forall \nu)$ whereas, by contrast, large oscillations in $[-A \omega ; A \omega]$ are observed. 

In order to understand the physical origin of these oscillations, let us first consider the equations governing the dynamics (Sec.~\ref{sec:principle})
in the limit of large accelerations~$\Gamma$. 
In this limit, the ball experiences large bounces so that,
since its potential energy at the collisions is negligible compared to its kinetic energy, 
we have $u_\mathrm{n}^\mathrm{+} \simeq - u_\mathrm{n+1}^\mathrm{-} \equiv \nu_\mathrm{n+1}$. 
Then, introducing the phase at impact~n, $\Phi_\mathrm{n} \equiv \omega t_\mathrm{n}$,
and remembering that $v_\mathrm{n}= - A \omega ~\sin(\omega t_\mathrm{n})$, one can write from Eqs.~(\ref{eq:restitution}) and (\ref{next_velocity}):
\begin{align}
\nu_\mathrm{n+1} &= e~\nu_\mathrm{n} - (1+e) ~A \omega ~\sin(\Phi_\mathrm{n}) \label{eq:recursion_nu}\\
\Phi_\mathrm{n+1} &= \Phi_\mathrm{n}+\frac{2 \omega}{g}\nu_\mathrm{n} \quad [2 \pi] \label{eq:recursion_phi}
\end{align}
Then, it is convenient to consider the limit of perfectly inelastic collisions, i.e. $e \rightarrow 0$.
The condition is obviously not satisfied experimentally but in this case, whatever the incoming velocity of the ball,
it bounces back with the velocity of the plate and: 
\begin{align}
\nu_\mathrm{n+1} &= - A \omega ~\sin(\Phi_\mathrm{n}) \label{eq:recursion_nu_bis} \\
\Phi_\mathrm{n+1} &= \Phi_\mathrm{n}-2 \Gamma \sin(\Phi_\mathrm{n}) \quad [2 \pi]
\label{eq:recursion_phi_bis}
\end{align}
Let us impose the impact velocity $\nu_\mathrm{n}$ at step n.
The relation $\nu_\mathrm{n}= - A \omega ~\sin(\Phi_\mathrm{n})$ implies that 
$\Phi_\mathrm{n-1}$ is fixed modulo $\pi$.
At step n, using Eq.~(\ref{eq:recursion_phi_bis}), these two values of $\Phi_\mathrm{n-1}$ result in two possible phases $\Phi_\mathrm{n}$:
\begin{align}
\Phi_\mathrm{n}^{(1)} &= - \arcsin \Bigl(\frac{\nu_\mathrm{n}}{A\omega} \Bigr) + \frac{2\Gamma}{A \omega} \nu_\mathrm{n}  \label{eq:recursion_phi_1} \\
\Phi_\mathrm{n}^{(2)} &= -\pi+ \arcsin \Bigl(\frac{\nu_\mathrm{n}}{A\omega} \Bigr) + \frac{2\Gamma}{A \omega} \nu_\mathrm{n}
\label{eq:recursion_phi_2}
\end{align}
Assuming that in the stationary state, both values of $\sin(\Phi_\mathrm{n})$ are equiprobable, one can estimate that the average plate velocity
$\overline{v}(\nu) = - \frac{1}{2}\,A\omega\,[\sin(\Phi_\mathrm{n}^{(1)})+\sin(\Phi_\mathrm{n}^{(2)})]$ and, after some simple algebra:
\begin{equation}
\frac{\overline{v}}{A\omega}=\frac{\pi}{\Gamma} \frac{\nu \omega}{\pi g} \cos \Bigl( 2 \pi ~\frac{\nu \omega}{\pi g}\Bigr), \quad -\frac{\Gamma}{\pi}<\frac{\nu \omega}{\pi g}<\frac{\Gamma}{\pi}
\label{eq:mean_v}
\end{equation}

\noindent
One sees that Eq.~(\ref{eq:mean_v}) nicely reproduces, at least qualitatively, the oscillations with linearly increasing amplitude that are observed experimentally (Fig.~\ref{fig:va_vp_S}). Moreover, the model accounts qualitatively for the period of the oscillations, $\frac{\pi g}{\omega}$, which is equal to the takeoff velocity
in the first periodic mode (Sec.~\ref{sec:restitution}).
However, Eq.~(\ref{eq:mean_v}) fails in accounting for the amplitude of the oscillations.
An almost quantitative agreement with the experimental amplitude (corresponding to $e = 0.95$ and $\Gamma = 2.35$) is obtained for $\Gamma \simeq 35$,
which is not surprising because the perfectly inelastic collisions ($e=0$) considered in our simplistic model impose strong vibrations to compensate the energy losses.
The model obviously suffers from several flaws. Taking $e=0$, we assumed that the ball takes off with the velocity of the plate, which has two consequences.
First, values of $\nu$ cannot be larger than $A\omega$. Second, even worse, in this peculiar limit the trajectory is necessarily periodic.
We avoided the problem by considering anyway that the trajectory is chaotic and, thus, rather the limit $e \ll 1$ such that the term $e~\nu_\mathrm{n}$ is negligible in Eq.~(\ref{eq:recursion_nu}). We expect the results to remain qualitatively valid at least in the limit of small restitution coefficient.

In conclusion, the ball and plate motions exhibit strong and complex correlations which are, at least partly, due to the synchronization
to the underlying periodic trajectories, as proven by the oscillations in Fig.~\ref{fig:va_vp_S}. In next section~\ref{sec:aperiodic},
we describe how our original device can be used to get rid of the synchronization, which, in turn, makes it possible to study
the remaining correlations and to assess, both experimentally and theoretically, the average energy of the ball in the chaotic regime.

\subsection{Aperiodic vibrations}
\label{sec:aperiodic}

\subsubsection{Plate motion}
\label{sec:platemotion}

In our experimental device, the amplifier connected to the permanent magnet shaker is fed by the signal
from the data acquisition board which has two analog outputs. In principle, this signal can be arbitrary.
However, because of the mechanical response of the shaker, one must carefully chose the waveform and, 
in particular, avoid discontinuities.

By construction, the shaker can be considered as a spring-mass system with losses,
driven by a force proportional to the input signal $s(t)$.
The equation of motion can be written in the form:
\begin{equation}
M  \ddot{z}(t) + \gamma \dot{z}(t) + k z(t) = \alpha\,s(t)
\label{eq:response}
\end{equation}
where $M$ stands for a mass, $k$ for a stiffness, $\alpha$ a proportionality coefficient and $\gamma$ a frictional coefficient which accounts for the losses.

The motion of the plate is thus governed by a second order differential equation.
If one manages to drive the system with a signal $s(t)$ which insures the continuity of all the derivatives up to the second order,
the shaker follows smoothly the driving signal. This can be achieved by building a signal $s(t)$ composed from a series
of sine cycles $s_\mathrm{i}(t) = a_\mathrm{i}\,\sin(\omega_\mathrm{i}\,t)$ insuring the continuity of $z(t)$, $\dot z(t)$ and $\ddot z(t)$.

To do so, we built the signal as follows:
at each cycle $\mathrm{i}$, the frequency $f_\mathrm{i}$ is randomly chosen in the interval $[35;45]$~Hz.
The period $i$ begins with $s = 0$ for a duration $1/f_\mathrm{i}$, which insures that $s$ is again zero at the end.
Notice that this automatically insures that $\ddot s = 0$ at both start and end.
Only $a_\mathrm{i}$ must be chosen with care so as to insure the continuity of the velocity. 
From the transfer function $\tilde{H}(\omega) \equiv {\tilde{z}}/{\tilde{s}}$ of the mechanical system
(easily obtained by measuring the response of the system to sinusoidal signals),
we estimate the amplitude $a_\mathrm{i}$ that insures the continuity of $A_\mathrm{i}\,\omega_\mathrm{i}$.

Doing so, we produce an aperiodic vibration of the plate which exhibits however a well-defined value of the velocity $A\omega$.

\begin{figure}[t]
\begin{center}
\includegraphics[width=0.8\columnwidth]{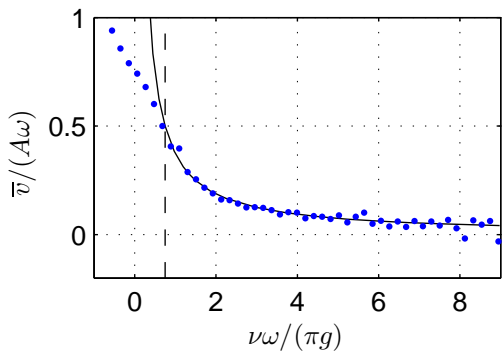}
\end{center}
\caption{\label{fig:va_vp_MF} Averaged ball velocity $\overline{v}$ vs. incoming velocity $\nu$ with an aperiodic vibration --
We display the results from an experimental trajectory (blue dots) and the theoretical prediction [black continuous line, Eq.~(\ref{mean_v})].
This prediction is only valid for $\nu > A \omega$,
this limit being represented by the vertical dashed line
[Experimental results: $f=40$ Hz, $\Gamma=2.35$].}
\end{figure}

\subsubsection{Ball-plate correlations}
\label{sec:correlations_FM}

Getting rid of the synchronization, we observe the disappearance of the oscillations when $\overline{v}$ is displayed against $\nu$ (Fig.~\ref{fig:va_vp_MF}).
However, the ball and plate motions remain correlated as proven by the nonzero $\overline{v}$ and its decreases with increasing $\nu$ .

This behavior can be accounted for in the following way.
Let us consider that the ball approaches the plate and that, at $t_0$, its altitude $h(t_0)=A$,
the amplitude of the plate motion in the corresponding cycle.
Let us further assume that the acceleration $A\,\omega^2$ is large enough,
so that one can neglect the variation of the ball velocity in the plate region ($h \in [-A,+A]$).
The ball-plate distance at time $t$ is $D(t) = A - \nu\,(t-t_0) - A\,\cos(\omega\,t)$.
The time $t_0$ accounts for the phase of the ball motion with respect to the plate motion.

First, one can determine how the collision time $t_c$, defined by $D(t_c)=0$, depends on $t_0$.
By differentiation of the condition $D(t_c)=0$ with respect to $t_0$ and $t_c$ and using $v(t)=- A\,\omega\,\sin(\omega\,t)$,
one gets $dt_0 = \frac{\nu + v(t_c)}{\nu}\,dt_c$.
Then, assuming that all $\Phi_0 \equiv \omega t_0$ have the same probability, with the density $\frac{1}{2\pi}$
(phase average, i.e. the ball and the plate motions are not synchronized),
the probability that the ball has an incoming velocity $\nu$ strikes the plate with the phase $\Phi \equiv \omega\,t_c$ is: 
\begin{equation}
P_{\nu}(\Phi)=\frac{1}{2 \pi} \frac{\nu + A\,\omega\,\sin(\Phi)}{\nu}
\label{mean_over_phi}
\end{equation}
Using this result, one can calculate the average of any quantity of interest, $x$, under the assumption
that the ball has a random initial phase $\Phi_0$, by writing
\begin{equation}
\overline{x} \bigr|_{\nu} = \int x~P_{\nu}(x) ~\mathrm d x =\int_0^{2 \pi} x ~P_{\nu}(\Phi) ~\mathrm d \Phi .
\label{mean_x_over_phi}
\end{equation}
In particular, applied to the plate velocity, $v = - A\,\omega\,\sin(\Phi)$, Eq.~(\ref{mean_x_over_phi}) leads to:
\begin{equation}
\overline{v} = - \int_0^{2 \pi}\,A\,\omega\,\sin(\Phi)\,P_{\nu}(\Phi)~\,d\Phi = \frac{(A\,\omega)^2}{2 \nu}
\label{mean_v}
\end{equation}
is in excellent agreement with the experimental data for $\nu > A\,\omega$ (Fig.~\ref{fig:va_vp_MF}).

The evolution of $\overline{v}(\nu)$ for $\nu \in [-A \omega ; A \omega]$ (note that, even if less probable, collisions are possible for a ball moving upwards)
is more difficult to account for analytically.
Indeed, in this case, there exists a forbidden region (a range of plate velocities or, equivalently, of phases) in which the ball cannot touch the plate.
In the case of such strict shadow effect,
the physical origin of the decrease of $\overline{v}(\nu)$ when $\nu$ increases remains the same but the solution of the problem is more complicate
and we shall not discuss it herein.
Indeed, in the chaotic regime, the relative acceleration being large, the ball bounces energetically most of the time 
and impinges generally onto the plate with a large velocity $\nu$.
Small velocity events are scarce and, in what follows, we will neglect their contribution.

In conclusion, when an aperiodic vibration of the plate is considered, the only remaining correlation between 
the ball and plate motions originates from a shadow effect. The ball more probably touches an ascending
than a descending plate. As a result, in average, the plate velocity $\overline{v}$ seen by the ball
increases when the impact velocity $\nu$ decreases and approaches the maximum plate velocity $A\,\omega$.
In the other limit, $\overline{v} \to 0$ for $\nu \to \infty$. 
As a result, the ball having a large kinetic energy sees a plate at rest and loses energy because of the collision
whereas the ball having a small kinetic energy is kicked by the plate and gains energy. This is the reason why the system
reaches a permanent regime, corresponding to a given, finite, average energy. 

\subsubsection{Energy}
\label{sec:energy of the bead}

The kinetic energy of the ball at the collisions, ${\mathcal E} = \frac{1}{2} m \nu^2$, is a major feature of its dynamics in the chaotic regime.
Measuring and predicting its average as a function of the characteristics of the plate motion
is not trivial and has been the subject of several studies.\cite{Warr1995,Geminard2003}
We first discuss the problem theoretically, using the conclusions of previous Sec.~\ref{sec:correlations_FM}, and then compare with experimental results.

In the following, we denote $\langle . \rangle$ the statistical average
over the N impacts experimentally obtained: $\langle x \rangle=\frac{1}{N}\sum_{n=1}^N x_\mathrm{n}$
or, for continuous variables, $\langle x \rangle=\int P(x)~\mathrm d x$. The probability density function of $x$, $P(x)$,
can also be written as a function of its conditional probability $P_{\nu}(x)$: $P(x)=\int P_{\nu}(x) P(\nu) ~\mathrm d \nu$.
Thus, using Eqs.~(\ref{mean_over_phi}) and~(\ref{mean_x_over_phi}), one gets:
\begin{equation}
\langle x \rangle=\int \overline{x} \bigr|_{\nu} P(\nu) ~\mathrm d \nu
\label{mean_x}
\end{equation} 

\begin{figure}[t]
\begin{center}
\includegraphics[width=0.8\columnwidth]{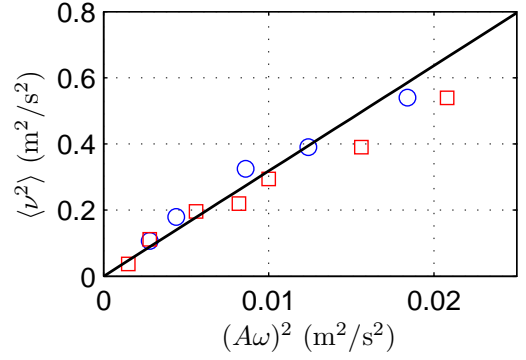}
\end{center}
\caption{\label{fig:energy} Mean square velocity $\langle \nu^2 \rangle$ vs. $(A \omega)^2$ --
The agreement is excellent for an aperiodic vibration (squares) and for a sinusoidal vibration (circles).
The black line corresponds to Eq.~(\ref{eq:mean_u2}) with $e=0.94$
[$\Gamma \in [1.34;4.40]$. Aperiodic: $f \in [35;45]$ Hz. Sine: $f=40$ Hz].}
\end{figure}

Let us now consider the energy of the ball around collision~n. Taking the square of the collision law Eq.~(\ref{eq:restitution}), we have:
\begin{equation}
(u_\mathrm{n}^\mathrm{+})^2=e^2(u_\mathrm{n}^\mathrm{-})^2- 2e(1+e)u_\mathrm{n}^\mathrm{-}v_\mathrm{n}+(1+e)^2 (v_\mathrm{n})^2
\label{u2}
\end{equation} 
All these quantities are linked with impact n, so that one can consider their average over the phase, $\Phi_0$,
of the incoming ball for a given value of the incoming velocity~$\nu$.
Using Eqs.~(\ref{mean_over_phi}) and~(\ref{mean_x_over_phi}) (remember that $\nu \equiv -u^-$):
\begin{align}
\overline{u_\mathrm{}^\mathrm{-} v}\Bigr|_{\nu} &= - \int_0^{2 \pi} A\,\omega\,\sin(\Phi)\,u_\mathrm{}^\mathrm{-}\,P_{\nu}(\Phi)\,d\Phi= - \frac{(A\,\omega)^2}{2}\\
\overline{v_\mathrm{}^2}\Bigr|_{\nu} &= \int_0^{2 \pi} (A\,\omega)^2\,\sin^2(\Phi)\,P_{\nu}(\Phi)\,d\Phi = \frac{(A\,\omega)^2}{2}.
\end{align}
Thus, the phase-average of Eq.~(\ref{u2}) for a given value of $\nu$ (the average is independent of the collision index $\mathrm n$) gives:
\begin{equation}
\overline{(u_\mathrm{}^\mathrm{+})^2}\Bigr|_{\nu} = e^2 \overline{(u_\mathrm{}^\mathrm{-})^2}\Bigr|_{\nu} + (1+e)(1+3e)\frac{(A\,\omega)^2}{2}.
\label{eq:u2_proba}
\end{equation}
Applying Eq.~(\ref{mean_x}) to Eq.~(\ref{eq:u2_proba}), one gets:
\begin{equation}
\langle (u^+)^2 \rangle=e^2~\langle (u^-)^2 \rangle+ (1+e)(1+3e)\frac{(A\,\omega)^2}{2}
\label{u2_bis}
\end{equation}
In addition, the energy balance immediately after impact n and before impact n+1 leads to
$(u_\mathrm{n}^\mathrm{+})^2+2 g z_\mathrm{n}=(u_\mathrm{n+1}^\mathrm{-})^2+2 g z_\mathrm{n+1}$.
Noticing that the potential energy of the ball at the collisions is unchanged, in average,
during the motion [i.e. $\langle z_\mathrm{n+1}-z_\mathrm{n} \rangle=0$],
one has $\langle (u_\mathrm{}^\mathrm{-})^2 \rangle=\langle (u_\mathrm{}^\mathrm{+})^2 \rangle\equiv \langle \nu^2 \rangle$.
In conclusion Eq.~(\ref{u2_bis}) can be written:
\begin{equation}
\langle \nu^2 \rangle=\frac{1 + 3 e}{1 - e}\frac{(A \omega)^2}{2},
\label{eq:mean_u2}
\end{equation} 
in excellent agreement with the experimental data (Fig.~\ref{fig:energy}).

We remind ourselves here that we made the assumptions that the ball was entering the plate region with a completely random phase
and that the contribution of the small impact velocities was negligible due to the scarcity of the events.
First, the use of an aperiodic vibration avoids the synchronization of the ball trajectory with that of the plate.
Second, the agreement of Eq.~(\ref{eq:mean_u2}) with the experimental data indicates that the assumption the small energy impacts are scarce is satisfied.
We mention that the result only holds true as long as the small energy events can be neglected,
thus at large enough accelerations. When the acceleration is decreased, one measures that the ball energy is smaller than expected from Eq.~(\ref{eq:mean_u2})
and even transitions to periodic motion.\cite{Geminard2003}

We point out that Eq.~(\ref{eq:mean_u2}) corresponds to a statistical average over the collisions.
It is thus not the energy an experimentalist would measure by taking values of the energy, at random times, along the ball trajectory.
Indeed, the duration of the flight after collision n, $\Delta t_\mathrm{n}$ scales like $\sqrt{\mathcal E_\mathrm{n}}$ (Eq.~\ref{next_velocity}).
As a consequence, he would measure $\langle \mathcal E \rangle_t = \frac{1}{2} m \langle \nu^2 \rangle_t$, where: 
\begin{equation}
\langle \nu^2 \rangle_t \equiv \frac{\sum_\mathrm{n} \nu_\mathrm{n}^2~\Delta t_\mathrm{n}}{\sum_\mathrm{n} \Delta t_\mathrm{n}}
\label{eq:mean_u2_temp_def}
\end{equation}
is the time average of the squared velocity.

Using results of Warr \textit{et. al.}, we show that, in the chaotic regime at large acceleration $\Gamma$, the two average energies are proportional.\cite{Warr1996}
Indeed, using a continuous description, one can write the two averages in the form:
\begin{align}
\langle \mathcal{E} \rangle &= \frac{\int_0^\infty \mathcal{E} P(\mathcal{E}) ~\mathrm d \mathcal{E}}{\int_0^\infty P(\mathcal{E}) ~\mathrm d \mathcal{E}} 
\label{eq:mean_E} \\
\langle \mathcal{E} \rangle_t &= \frac{\int_0^\infty \mathcal{E} \Delta t P(\mathcal{E}) ~\mathrm d \mathcal{E}}{\int_0^\infty \Delta t P(\mathcal E) ~\mathrm d \mathcal{E}}
\label{eq:mean_E_temp}
\end{align}
where $P(\mathcal E)$ denotes the probability distribution function (PDF) of the energy of the ball during a flight between two collisions.
Warr \textit{et. al} proved that this PDF follows a Boltzmann law: $P(\mathcal E) \propto \exp(-\beta \mathcal E)$, which we also observe experimentally (not shown).
Using this result, integrating Eqs.~(\ref{eq:mean_E}) and~(\ref{eq:mean_E_temp}) by part, provided that $\Delta t \propto \sqrt{\mathcal E}$,
one obtains $\langle \mathcal{E} \rangle_t =\frac{3}{2}\langle \mathcal{E} \rangle$ or, equivalently:
\begin{equation}
\langle \nu^2 \rangle_t=\frac{3}{2}\langle \nu^2 \rangle=\frac{3}{2}\frac{1 + 3 e}{1 - e}\frac{(A \omega)^2}{2}.
\label{mean_u2_temp}
\end{equation} 
The time average is larger than the average over the collisions, which is due to the fact that larger energies are associated to larger durations of the flight.
Warr \textit{et. al},\cite{Warr1995} using the formalism of a discrete-time Langevin equation,\cite{Risken1984} obtained a similar result which is consistent
with Eq.~(\ref{mean_u2_temp}) in the limit $e \rightarrow 1$.
Thus, it is interesting to notice that, in this peculiar case, measuring the average energy over the discrete collisions is equivalent, to within a prefactor, to measuring the temporal average of the energy.

\section{Discussion and conclusion}
\label{sec:conclusion}

We reported on an experiment dedicated to the study of the trajectories of a ball bouncing randomly on a vibrating plate.
We showed how to measure the collision times and the plate position such that the entire trajectory of the ball can
be reconstructed. We first use the experimental device to measure the ball-plate restitution coefficient in three different ways.
Then, for a ball exhibiting a chaotic trajectory, we revealed the correlations between the ball and the plate motions
and showed how the use of an aperiodic motion makes it possible to avoid part of the correlations.
In this case, an analytic expression of the average plate velocity seen by the colliding ball can be obtained.
The result, in excellent agreement with the experimental measurements, makes it possible to propose an analytic expression for the average energy of the ball, again in excellent agreement with the experiments.
Interestingly, we observe that the agreement is also pretty good for the sinusoidal vibration, in spite of the underlying correlations between the ball and plate motions.


\end{document}